\documentclass[prd,aps,twocolumn,nofootinbib,showpacs]{revtex4-1}

\usepackage{amsmath,graphicx,color,epsfig}
\usepackage{pstricks}
\usepackage{float}
\usepackage{subfigure}
\usepackage{color}

\usepackage[dvipdfm, colorlinks=true, linkcolor=blue, citecolor=blue, urlcolor=blue] {hyperref}

\begin{document}

\title{General Properties on Applying the Principle of Minimum Sensitivity to High-order Perturbative QCD Predictions}

\author{Yang Ma}
\author{Xing-Gang Wu}
\email{email:wuxg@cqu.edu.cn}
\author{Hong-Hao Ma}
\author{Hua-Yong Han}

\address{Department of Physics, Chongqing University, Chongqing 401331, P.R. China}
\address{Institute of Theoretical Physics, Chongqing University, Chongqing 401331, P.R. China}

\date{\today}

\begin{abstract}

As one of the key components of perturbative QCD theory, it is helpful to find a systematic and reliable way to set the renormalization scale for a high-energy process. The conventional treatment is to take a typical momentum as the renormalization scale, which assigns an arbitrary range and an arbitrary systematic error to pQCD predictions, leading to the well-known renormalization scheme and scale ambiguities. As a practical solution for such scale setting problem, the ``Principle of Minimum Sensitivity'' (PMS), has been proposed in the literature. The PMS suggests to determine an optimal scale for the pQCD approximant of an observable by requiring its slope over the scheme and scale changes to vanish. In the paper, we present a detailed discussion on general properties of PMS by utilizing three quantities $R_{e^+e^-}$, $R_\tau$ and $\Gamma(H\rightarrow b\bar{b})$ up to four-loop QCD corrections. After applying the PMS, the accuracy of pQCD prediction, the pQCD convergence, the pQCD predictive power and etc., have been discussed. Furthermore, we compare PMS with another fundamental scale setting approach, i.e. the Principle of Maximum Conformality (PMC). The PMC is theoretically sound, which follows the renormalization group equation to determine the running behavior of coupling constant and satisfies the standard renormalization group invariance. Our results show that PMS does provide a practical way to set the effective scale for high-energy process, and the PMS prediction agrees with the PMC one by including enough high-order QCD corrections, both of which shall be more accurate than the prediction under the conventional scale setting. However, the PMS pQCD convergence is an accidental, which usually fails to achieve a correct prediction of unknown high-order contributions with next-to-leading order QCD correction only, i.e. it is always far from the ``true" values predicted by including more high-order contributions.


\end{abstract}

\maketitle

\section{Introduction}
\label{sec:int}

According to the renormalization group (RG) invariance~\cite{peter,rgi1,rgi2,callan,symanzik,peter2}, a physical observable should not depend on any ``unphysical'' choices. In another words, the RG-invariance indicates that the dependence of an observable on the renormalization scheme and scale should vanish. However, for fixed-order pQCD approximations, the renormalization scheme and scale dependence from both the running coupling and the corresponding expansion coefficients at the same order do not exactly cancel. To deal with a fixed-order calculation, one usually takes the renormalization scale as the typical momentum transfer of the process, or a value to minimize the contributions of large loop diagrams, and varies it over a certain range to ascertain its uncertainty. This conventional scale setting procedure leads to well-known renormalization scheme and scale ambiguities and assigns an arbitrary range and an arbitrary systematic error to fixed-order pQCD predictions. To solve such renormalization scheme and scale ambiguities, it is helpful to find a general way to set the optimal scale and hence the optimal running behavior of strong coupling constant for any processes via a process-independent and systematic way, cf. a recent review on QCD scale setting~\cite{pmcreview}.

To compare with the conventional scale setting, it has been suggested by Stevenson at 1981 that one can achieve a good prediction for an observable by requiring its pQCD approximant to be minimum sensitive to the variations of those unphysical parameters. This treatment is called as the ``Principle of Minimum Sensitivity (PMS)"~\cite{PMS,PMS3,PMS4}. The PMS admits that different scheme and scale choices do lead to theoretical uncertainties, however the ``true" prediction of an observable can only be achieved by using optimal scheme and scale. The scheme dependence of the PMS predictions have been analyzed in Refs.\cite{fixed3,fixed4}. It is noted that the PMS satisfies local RG-invariance~\cite{pmcma}, which provides a practical approach to systematically fix the optimal scheme and scale for high-energy process. It has been noted that after applying the PMS, the pQCD prediction does show a fast steady behavior over the scheme and scale changes. As an example, it has been applied to study the fixed-point behavior of the coupling constant at the low-energy region~\cite{fixed,fixed2}.

On the other hand, it has also been observed that PMS does not satisfy the RG-properties such as symmetry, reflexivity, and transitivity~\cite{pmccolloquium}. So the relations among different physical observables depend on the choice of intermediate renormalization scheme, leading to residual scheme dependence. Moreover, the predicted PMS scale for three-jet production via $e^+ e^-$-annihilation can not yield correct physical behavior at the next-to-leading order (NLO) level, i.e. it anomalously rises without bound for small jet energy~\cite{jet1,jet2}. There are even doubts on the usefulness of PMS~\cite{antisetting}. All those discussions indicate the necessity of further careful studies on theoretical principles underlying the PMS and on applications to more high-loop examples.

Great improvements on understanding the PMS procedures and on applying PMS scale setting to higher perturbative orders other than the NLO level have recently been achieved in Ref.\cite{PMS2}. In recent years, there are many progresses on studying the two-loop and higher QCD corrections. For examples, the quantities $R_{e^+e^-}$, $R_{\tau}$ and $\Gamma(H\to b\bar{b})$ have been calculated up to four-loop level under the $\overline{\rm MS}$-scheme~\cite{Ralphas1,Ralphas2,cor9,cor10}. With all those developments, it is possible to make a detailed discussion on general properties of PMS, and to show to what degree it can be applied. For the purpose, we shall present the PMS predictions for $R_{e^+e^-}$, $R_{\tau}$ and $\Gamma(H\to b\bar{b})$ up to four-loop level. General PMS properties, such as the accuracy of the pQCD prediction, the convergence of the perturbative series, the predictive power of pQCD theory and etc, shall be discussed via comparing the predictions with those under the conventional scale setting.

Recently, another well-known scale setting approach, i.e. the Brodsky-Lepage-Mackenzie approach suggested by Brodsky etal. at 1983~\cite{BLM}, has been developed into a fundamental one, i.e. the ``Principle of Maximum Conformality (PMC)"~\cite{pmc1,pmc2,pmc3,pmc4,pmc5,BMW,BMW2}. In different to PMS scale setting, the PMC states that we should determine different optimal scales for the high-energy process under different schemes, and the final predictions are independent on the scheme choices due to commensurate scale relations~\cite{commensurate} and also the scheme-independence of a conformal series. The running behavior of the coupling constant is governed by the RG-equation~\cite{beta0,beta00,beta1,beta11,beta2,beta3,beta4}. Inversely, the PMC states the optimal behavior/scale of the coupling constant can be achieved by using the $\beta$-terms in perturbative series. The PMC follows standard RG-invariance and satisfies all RG-properties~\cite{pmccolloquium}. When one applies the PMC, the scales of the coupling constant are shifted at each order such that no contributions proportional to the QCD $\beta$-function remain. The resulting pQCD series is thus identical to a scheme-independent conformal series. Since the resulting series is free of divergent renormalon terms~\cite{renormalon,renormalon1}, the pQCD convergence can be naturally improved. The PMS and PMC scale settings have quite different starting points and their predictions usually have quite different perturbative nature, it is thus helpful to present a detailed comparison of the PMS predictions with the PMC ones.

The remaining parts of the paper are organized as follows. In Sec.\ref{sec:PMS} we first present a short review on local RG-invariance that underlies PMS, then, we present the PMS formulas up to high-perturbative orders. A tricky way to derive the PMS RG-invariants at high-orders in the Appendix. In Sec.\ref{sec:App} we investigate the PMS properties based on three quantities $R_{e^+e^-}$, $R_\tau$ and $\Gamma(H\rightarrow b\bar{b})$ up to four-loop level. In Sec.\ref{sec:vs} we present a detailed comparison of PMS and PMC via the quantity $R_{e^+e^-}$. Sec.\ref{sec:summary} is reserved for a summary.

\section{Calculation technology for the PMS scale setting}
\label{sec:PMS}

Conventionally, the running behavior of the strong coupling constant is controlled by the following $\beta^{\cal R}$-function or the RG-equation,
\begin{eqnarray}
\beta^{\cal R}=\mu^2\frac{\partial}{\partial\mu^2} \left(\frac{\alpha^{\cal R}_s(\mu)}{4\pi}\right) =-\sum_{i=0}^{\infty}\beta^{\cal R}_{i}\left(\frac{\alpha^{\cal R}_s(\mu)}{4\pi}\right)^{i+2}, \label{beta}
\end{eqnarray}
where $\mu$ stands for the renormalization scale, and the superscript $\cal R$ stands for an arbitrary renormalization scheme (usually taken as the $\overline{\rm MS}$-scheme). For convenience and without introducing any confusion, we shall omit the superscript ${\cal R}$ in the following formulas. The first two $\beta$-terms, $\beta_{0}= 11-\frac{2}{3}n_f$ and $\beta_{1}= 102-\frac{38}{3}n_f$, are scheme independent, where $n_f$ is the number of active flavors; while the $\beta_n$-terms with $\left(n\geq2\right)$ are scheme dependent~\cite{beta1,beta11,beta2,beta3,beta4}. The scheme dependence/transformation for high-order $\beta$-terms have been discussed in Refs.\cite{robert1,robert2,robert3}.

It is convenient to use $\tau=\ln(\mu^2/\tilde\Lambda^2_{\rm QCD})$ and $\beta_{n\geq2}$ to label a particular choice of renormalization scale and renormalization scheme~\cite{PMS}. Here $\tilde\Lambda_{\rm QCD}$ is the reduced asymptotic scale, which is defined as
\begin{eqnarray}
\tilde\Lambda_{\rm QCD}=\left(\frac{\beta_1} {\beta_0^2}\right)^{-\beta_1/2\beta_0^2} \Lambda_{\rm QCD} .
\end{eqnarray}
We can study the scale- and scheme- dependence of the pQCD predictions via the extended RG-equations~\cite{PMS,HJLu}.

\subsection{Local RG-invariance and PMS}

As an illustration of local RG-invariance, we deal with the perturbative approximant $(\varrho_n)$ for an arbitrary physical observable $\varrho$, which can be written as
\begin{eqnarray} \label{pmsstartpoint}
\varrho_n(Q) = {\cal C}_{0}(Q) a_s^p(\mu) + \sum_{i=1}^{n}{\cal C}_i(Q,\mu) a_s^{i+p}(\mu) , \label{phy}
\end{eqnarray}
where $Q$ is the experimental scale at which it is measured, $a_s=\alpha_s/\pi$, and $p$ is the power of coupling constant associated with tree-level term. The calculation of the coefficients ${\cal C}_i$ involves ultraviolet divergences which must be regulated and removed by a renormalization procedure. At the finite order, the pQCD predictions dependent on the choice of renormalization scheme and scale, i.e.
\begin{equation}\label{pmsfinite}
\partial \varrho_n /\partial {\rm (RS)}={\cal O}(a_s^{p+n}),
\end{equation}
where ${\rm RS}$ stands for the scheme or scale parameter, respectively.  Eq.(\ref{pmsfinite}) shows the self-consistency of a perturbation theory, i.e. the $\rm{N}^n$-LO approximate $\varrho_n$ must agree to ${\cal O}(a_s^{p+n})$ under different choices of scheme and scale. The tree-level coefficient ${\cal C}_0$ is scheme and scale independent, we set its value to be $1$ in later calculations. When ${\cal C}_0 \neq 1$, the results can be obtained via the transformation, ${\cal C}_i(Q,\mu) \to {\cal C}'_i(Q,\mu)={\cal C}_i(Q,\mu)/{\cal C}_0$.

As mentioned in the Introduction, there are renormalization scheme and scale ambiguities for the fixed-order pQCD approximant $\varrho_n$. The PMS suggests to eliminate such scheme and scale ambiguities by finding optimal scheme and optimal scale of the process, which can be achieved by requiring $\varrho_n$ to satisfy the following equations~\cite{PMS,PMS2},
\begin{eqnarray}
\frac{\partial \varrho_n}{\partial \tau} &=& 0,  \\
\frac{\partial \varrho_n}{\partial \beta_m} &=& 0. \qquad (m=2,...,n)
\end{eqnarray}
They can be further written as
\begin{widetext}
\begin{eqnarray}
\frac{\partial \varrho_n}{\partial \tau}&=&\left(\left.\frac{\partial }{\partial \tau}\right|_{a_s} +\beta (a_s) \frac{\partial}{\partial (a_s/4)} \right)\varrho_n=0 , \label{finite-sca} \\
\frac{\partial \varrho_n}{\partial \beta_m}&=&\left(\left.\frac{\partial }{\partial \beta_m}\right|_{a_s} -\beta (a_s)\int_0^{a_s/4} d\left(\frac{a_s^\prime}{4}\right) \frac{(a_s^\prime/4)^{m+2}}{\left[\beta (a_s^\prime)\right]^2} \frac{\partial} {\partial (a_s/4)} \right)\varrho_n=0 , \;\;\; (m=2,3,...) \label{finite-sch}
\end{eqnarray}
\end{widetext}
where the integration in the second equations can be treated via the $\alpha_s$-expansion,
\begin{widetext}
\begin{displaymath}
\beta (a_s)\int_0^{a_s/4} d\left(\frac{a_s^\prime}{4}\right) \frac{(a_s^\prime/4)^{m+2}}{\left[\beta (a_s^\prime)\right]^2} =-\frac{\left(a_{\rm s}/4\right)^{j+1}}{\beta_0} \left(\frac{1}{j-1}-\frac{\beta_1}{\beta_0} \frac{j-2}{j(j-1)} \left(\frac{a_{\rm s}}{4}\right)+\dots\right).
\end{displaymath}
\end{widetext}
The standard RG-invariance states that only the physical observable $\varrho=\varrho_n|_{n\to\infty}$ agrees with those equations. Thus, using Eqs.(\ref{finite-sca},\ref{finite-sch}) for the fixed-order approximant is theoretically unsound, and they instead introduce a kind of local RG-invariance~\cite{pmcma}. This provides the reason why the PMS does not satisfy the basic RG-properties~\cite{pmccolloquium}. The PMS, however, provides an intuitive way to set the optimal scheme and optimal scale, and its resultant tends to be steady over the scheme and scale changes around the optimal point.

The running behavior of strong coupling constant can be obtained via solving RG-equation (\ref{beta}), which can be rewritten as
\begin{eqnarray}
\tau &=& \int_{a_s/4}^\infty d \left( \frac{x}{4} \right) \frac{1}{\beta^{(n)}(x)} \nonumber\\
& =& \frac{4}{\beta_0 a_s}+\frac{\beta_1}{\beta_0^2}\ln \left| \frac{\beta_1 a_s}{\beta_1 a_s+4\beta_0}\right|+\Delta(a_s), \label{betaint}
\end{eqnarray}
where
\begin{equation}
\Delta(a_s)=\int_0^{a_s/4} d \left( \frac{x}{4} \right) \left(\frac{1} {\beta^{(n)}(x)}-\frac{1}{\beta^{(1)}(x)}\right).
\end{equation}
The symbol $\beta^{(n)}$ stands for the cut $\beta$-function up to $a_s^{n+2}$. Eq.(\ref{betaint}) is the ``integrated $\beta$-function equation'', or simply, the``int-$\beta$ equation'', which can be solved numerically.

In the following, we shall show how PMS applies local RG-invariance to set the optimal scale and how the RG-invariant coefficients at each order are derived.

\subsection{PMS procedures up to high-orders}

For a $\rm{N}^n$-LO pQCD approximate (\ref{pmsstartpoint}), we have to fix totally $2n+1$ variables for determining optimal scheme and optimal scale, i.e. $\tilde{a}_s$, $\tilde\tau$, $\tilde{\beta}_2$, $\cdots$, $\tilde{\beta}_n$, $\tilde{\cal C}_1$, $\cdots$, $\tilde{\cal C}_n$. Those parameters can be fixed by using $n$ local RG-equations (\ref{finite-sca},\ref{finite-sch}), one int-$\beta$ equation (\ref{betaint}), and also $n$ scheme-and-scale independent RG-invariants from the self-consistency relation (\ref{pmsfinite}). To be a useful reference for applying PMS scale setting, we take the QCD corrections up to $\rm{N}^3$-LO level as a detailed explanation.

At the NLO level, the NLO approximate is
\begin{displaymath}
\varrho_1 = a_s^p(1+{\cal C}_1 a_s).
\end{displaymath}
The NLO approximate $\varrho_1$ can be calculated in an initial choice of scheme (usually the $\overline{\rm MS}$-scheme) and scale. We have three parameters $\tilde{a}_s$, $\tilde{\tau}$ and $\tilde{\cal C}_1$ to be determined.

Differentiating $\varrho_1$ over $\tau$ and using the self-consistency relation (\ref{pmsfinite}), i.e. the coefficient at the order of ${\cal O}(a_s^{p+1})$ should be zero, we obtain
\begin{eqnarray}
{\partial {\cal C}_1 \over \partial \tau}= {1 \over 4}p \beta_0 . \label{part}
\end{eqnarray}
Integrating it over $\tau$, we get one RG-invariant integration constant $\rho_1$, which can be expressed as
\begin{equation}
\rho_1 = {1\over 4} p \beta_0 \tau-{\cal C}_{1} = {1\over 4} p \beta_0 \tilde{\tau}-\tilde{\cal C}_{1}, \label{rho1}
\end{equation}
where the second equation is from the RG invariance. As a tricky point, since $\rho_1$ depends solitarily on $Q$ at which the observable is measured, one can transform $\varrho_n(Q)$ as $\varrho_{n}(\rho_1)$. The advantage of such transformation lies in that, $\varrho_n(\rho_1)$ does not depend on $\Lambda_{\rm{QCD}}$, thus avoiding the uncertainties from the choice of $\Lambda_{\rm{QCD}}$.

From Eq.(\ref{finite-sca}), we obtain the NLO local RG-equation
\begin{eqnarray}
p \beta_0-\big[p \tilde a_s^{p-1}+(p+1)\tilde {\cal C}_1 \tilde a_s^p\big] \left(\beta_0+ \frac{\beta_1 \tilde a_s}{4}\right) =0, \label{NLOPMS}
\end{eqnarray}
which leads to
\begin{eqnarray}
\tilde {\cal C}_1=-\frac{ p\beta_1}{(p+1)(4\beta_0+\beta_1 \tilde a_s)}. \label{NLOr1}
\end{eqnarray}
Together with the NLO int-$\beta$ equation
\begin{eqnarray}
\tilde{\tau}=\frac{4}{\beta_0 \tilde{a}_s}+\frac{\beta_1}{\beta_0^2}\ln \left| \frac{\beta_1 \tilde{a}_s}{\beta_1 \tilde{a}_s +4\beta_0}\right|, \label{NLOtau}
\end{eqnarray}
we finally obtain
\begin{eqnarray}
\frac{1}{\tilde a_s}+\frac{ p\beta_1}{(p+1)(4\beta_0+\beta_1 \tilde a_s)}+\frac{\beta_1}{4\beta_0} \ln\left| \frac{\beta_1\tilde{a}_s}{\beta_1 \tilde a_s+4\beta_0}\right| = \rho_1 \label{NLOPMSfinal}.
\end{eqnarray}
From those equations (\ref{NLOr1},\ref{NLOtau},\ref{NLOPMSfinal}), we can derive $\tilde{\tau}$, $\tilde {\cal C}_1$, $\tilde{a}_s$, and finally the get optimized prediction for $\varrho_1$.

For high-order QCD corrections, we can apply similar procedures via a step-by-step way for determining all the parameters.

Using the self-consistency condition (\ref{pmsfinite}), the local RG invariants $\rho_n$ can be determined via an order-by-order way. Once a $\rho_n$ has been determined at a particular perturbative order, it shall be fixed for all high-order PMS treatment. Except for those local RG invariants, all other parameters should be re-determined when new high-order corrections are included.

At the N$^2$-LO level, we have five parameters to be determined, i.e. $\tilde{a}_s$, $\tilde{\tau}$, $\tilde{\beta}_2$, $\tilde{\cal C}_1$, and $\tilde{\cal C}_2$. There are two equations that can be obtained from the local RG-equations ${\partial \varrho_2}/{\partial \tau}=0$ and ${\partial \varrho_2}/{\partial \beta_2}=0$:
\begin{widetext}
\begin{eqnarray}
16 (2+p) \tilde{\cal C}_2 \beta_0+4 \left[(1+p) \tilde{\cal C}_1+(2+p) \tilde{\cal C}_2 \tilde{a}_s\right] \beta_1+\left[\tilde{a}_s \left(\tilde{\cal C}_1+2 \tilde{\cal C}_2 \tilde{a}_s\right)+p \left(1+\tilde{\cal C}_1 \tilde{a}_s+\tilde{\cal C}_2 \tilde{a}_s^2\right)\right] \tilde{\beta}_2 &=& 0 , \label{NNLOP1} \\
48 \left[(1+p) \tilde{\cal C}_1+(2+p) \tilde{\cal C}_2 \tilde{a}_s\right] \beta_0+\tilde{a}_s \left[\tilde{a}_s \left(\tilde{\cal C}_1+2 \tilde{\cal C}_2 \tilde{a}_s\right)+p \left(1+\tilde{\cal C}_1 \tilde{a}_s+\tilde{\cal C}_2 \tilde{a}_s^2\right)\right] \tilde{\beta}_2 &=& 0 .  \label{NNLOP2}
\end{eqnarray}
\end{widetext}

At the N$^3$-LO level, we have seven parameters to be determined, i.e. $\tilde{a}_s$, $\tilde{\tau}$, $\tilde{\beta}_2$, $\tilde{\beta}_3$, $\tilde{\cal C}_1$, $\tilde{\cal C}_2$, and $\tilde{\cal C}_3$. There are three local RG-equations that can be obtained from ${\partial \varrho_3}/{\partial \tau}=0$, ${\partial \varrho_3}/{\partial \beta_2}=0$, and ${\partial \varrho_3}/{\partial \beta_3}=0$:
\begin{widetext}
\begin{eqnarray}
&&64 (3+p) \tilde{\cal C}_3 \beta _0+16 \left((2+p) \tilde{\cal C}_2+(3+p) \tilde{\cal C}_3 \tilde{a}_s\right) \beta _1+4 \left((1+p) \tilde{\cal C}_1+ \tilde{a}_s \left((2+p) \tilde{\cal C}_2+(3+p) \tilde{\cal C}_3 \tilde{a}_s\right)\right) \tilde{\beta}_2  \nonumber\\
&&\quad\quad +\left(p+\tilde{\cal C}_1 \tilde{a}_s+p \tilde{\cal C}_1 \tilde{a}_s+2 \tilde{\cal C}_2 \tilde{a}_s^2+p \tilde{\cal C}_2 \tilde{a}_s^2+(3+p) \tilde{\cal C}_3 \tilde{a}_s^3\right) \tilde{\beta}_3=0 , \label{NNNLOP1}\\
&&384 \left[(2+p) \tilde{\cal C}_2+(3+p) \tilde{\cal C}_3 \tilde{a}_s\right] \beta_0^2-\tilde{a}_s \left\{p \left(1+\tilde{\cal C}_1 \tilde{a}_s+\tilde{\cal C}_2 \tilde{a}_s^2+\tilde{\cal C}_3 \tilde{a}_s^3\right)+\tilde{a}_s \left[\tilde{\cal C}_1+\tilde{a}_s \left(2 \tilde{\cal C}_2+3 \tilde{\cal C}_3 \tilde{a}_s\right)\right]\right\} \beta _1 \tilde{\beta}_2 \nonumber\\
&&\quad\quad +\left\{p \left(1+\tilde{\cal C}_1 \tilde{a}_s+\tilde{\cal C}_2 \tilde{a}_s^2+\tilde{\cal C}_3 \tilde{a}_s^3\right)+\tilde{a}_s \left[\tilde{\cal C}_1+\tilde{a}_s \left(2 \tilde{\cal C}_2+3 \tilde{\cal C}_3 \tilde{a}_s\right)\right]\right\} \beta _0 \left(8 \tilde{\beta}_2+3 \tilde{a}_s \tilde{\beta}_3\right)=0, \label{NNNLOP2} \\
&&96 \left\{(1+p) \tilde{\cal C}_1+\tilde{a}_s \left[(2+p) \tilde{\cal C}_2+(3+p) \tilde{\cal C}_3 \tilde{a}_s\right]\right\} \beta_0^2-8 \left\{p \left(1+\tilde{\cal C}_1 \tilde{a}_s+\tilde{\cal C}_2 \tilde{a}_s^2+\tilde{\cal C}_3 \tilde{a}_s^3\right)+\tilde{a}_s \left[\tilde{\cal C}_1+\tilde{a}_s \left(2 \tilde{\cal C}_2+3 \tilde{\cal C}_3 \tilde{a}_s\right)\right]\right\} \beta_0 \beta_1, \nonumber\\
&&\quad\quad +\tilde{a}_s \left\{p \left(1+\tilde{\cal C}_1 \tilde{a}_s+\tilde{\cal C}_2 \tilde{a}_s^2+\tilde{\cal C}_3 \tilde{a}_s^3\right)+\tilde{a}_s \left[\tilde{\cal C}_1+\tilde{a}_s \left(2 \tilde{\cal C}_2+3 \tilde{\cal C}_3 \tilde{a}_s\right)\right]\right\} \beta_1^2=0 . \label{NNNLOP3}
\end{eqnarray}
\end{widetext}

Up to ${\rm N}^3$-LO level, in addition to $\rho_1$, we need to determine two extra RG invariants $\rho_2$ and $\rho_3$, which can be fixed via a similar way as the NLO case, detailed procedures can be found in Refs.\cite{PMS2,pmcma,PMSrho}. Then, we obtain
\begin{eqnarray}
\rho_2 &=& {\cal C}_2-\frac{(1+p) {\cal C}_1^2}{2 p}-\frac{\beta _1 {\cal C}_1}{4 \beta _0}+\frac{p \beta _2}{16 \beta _0} \label{rho21} \\
&=& \tilde{\cal C}_2-\frac{(1+p) \tilde{\cal C}_1^2}{2 p}-\frac{\beta _1 \tilde{\cal C}_1}{4 \beta _0}+\frac{p \tilde{\beta}_2}{16 \beta _0}  \label{rho22}
\end{eqnarray}
and
\begin{widetext}
\begin{eqnarray}
\rho_3 &=& 2 {\cal C}_3+\frac{{\cal C}_1^2 \beta _1}{4 p \beta _0}-\frac{{\cal C}_1 \beta _2}{8 \beta _0}+\frac{p \beta _3}{64 \beta _0}  +\frac{2 (1+p) (2+p) {\cal C}_1^3}{3 p^2}-\frac{2 (2+p) {\cal C}_1 {\cal C}_2}{p} \label{rho31} \\
&=& 2 \tilde{\cal C}_3+\frac{\tilde{\cal C}_1^2 \beta _1}{4 p \beta _0}-\frac{\tilde{\cal C}_1 \tilde{\beta}_2}{8 \beta _0}+\frac{p \tilde{\beta}_3}{64 \beta_0}  +\frac{2 (1+p) (2+p) \tilde{\cal C}_1^3}{3 p^2}-\frac{2 (2+p) \tilde{\cal C}_1 \tilde{\cal C}_2}{p} . \label{rho32}
\end{eqnarray}
\end{widetext}
The first equations (\ref{rho21},\ref{rho31}) are to set the value of $\rho_{2,3}$ with the known parameters calculated under the initial scheme-and-scale choices, the second equations (\ref{rho22},\ref{rho32}) are due to scheme-and-scale independence of RG-invariants $\rho_{2,3}$. As a cross-check of those formulas, when setting $p=1$, we turn to the same expressions as those of Ref.\cite{PMS2,pmcma}.

As a summary, in combination with all local RG-equations, the known RG-invariants, and also the same order int-$\beta$ equation (\ref{betaint}), we are ready to derive all the wanted optimal parameters. This can be done numerically by following the ``spiraling'' method~\cite{pmcma,pmsfixedpoint1,pmsfixedpoint2}. For a general all-order determination, the procedures of the ``spiraling" method are
\begin{itemize}
\item Firstly, one takes an initial value for $\tilde{a}_s$, which can be approximated by using RG-equation at the same order at an arbitrary initial scale. This initial scale should be large enough to ensure the pQCD calculation, which can be practically (to short the number of iterations) taken as the typical momentum flow of the process.

\item Secondly, for the first iteration, one sets the initial values for the scheme-dependent $\tilde{\beta}_2$, $\cdots$, $\tilde{\beta}_n$ to be $\beta_2$, $\cdots$, $\beta_n$ that have been calculated under an initial renormalization scheme. For new iterations their values are replaced by the ones determined from the last iteration. Then, one solves the local RG-equations, similar to Eqs.(\ref{NLOPMS},\ref{NNLOP1},\ref{NNLOP2},\ref{NNNLOP1},\ref{NNNLOP2},\ref{NNNLOP3}), for $\tilde{\cal C}_1$, $\cdots$, $\tilde{\cal C}_n$.

\item Thirdly, one applies the calculated value of $\tilde{\cal C}_1$, $\cdots$, $\tilde{\cal C}_n$ into the equations on RG-invariants $\rho_1$, $\cdots$, $\rho_n$, similar to Eqs.(\ref{rho1},\ref{rho22},\ref{rho32}), for $\tilde{a}_s$, $\tilde{\tau}$, $\tilde{\beta}_2$, $\cdots$, $\tilde{\beta}_n$.

\item Finally, one iterates from second step until the results for $\varrho_n$ converge to an acceptable prediction.
\end{itemize}

As a remarkable feature of renormalization theory, even if the coefficients ${\cal C}_n$ and the $\beta$-terms $\beta_n$ are separately different in different schemes, there exist some combinations of them that are RG-invariant. The above derived integration parameters $\rho_n$ are such kind of RG-invariants, which are key components to determine the ``optimal $\varrho_n$''. Because $\rho_n$ are RG-invariants, one can demonstrate that the final PMS predictions are independent of any choice of initial scale, being consistent with one of requirement of basic RG-invariance~\cite{pmcreview}. Thus, to apply PMS, one can simply set the initial scale to be a typical one such as the typical momentum of the process or the one at which the observable is measured. This, inversely, provides us a simpler/tricky way to derive the RG-invariants $\rho_n$, which are put in the Appendix.

\section{General properties and Applications of PMS scale setting}
\label{sec:App}

In this section, we shall present a detailed discussion on general properties of PMS scale setting by utilizing three quantities $R_{e^+e^-}$, $R_\tau$ and $\Gamma(H\rightarrow b\bar{b})$ up to four-loop level. A comparison of PMS and conventional scale settings shall also be presented.

\subsection{$R_{e^+e^-}$ up to four-loop QCD corrections}
\label{sec:Ree}

The $e^+e^-$ annihilation provides one of the most precise tests of pQCD theory. Its measurable quantity, i.e. the $R$-ratio $R(Q)$, is defined as
\begin{eqnarray}
R_{e^+e^-}(Q)&=&\frac{\sigma\left(e^+e^-\rightarrow {\rm hadrons} \right)}{\sigma\left(e^+e^-\rightarrow \mu^+\mu^-\right)}\nonumber\\
&=& 3\sum_q e_q^2\left[1+R(Q)\right], \label{Re+e-}
\end{eqnarray}
where $Q$ stands for the $e^+e^-$ collision energy at which the $R$-ratio is measured. The pQCD approximant for $R(Q)$ up to $(n+1)$-loop correction can be written as
\begin{eqnarray}
R_n(Q,\mu_0)=\sum_{i=0}^{n} {\cal C}_{i}(Q,\mu_0) a_s^{i+1}(\mu_0), \label{R(Q)}
\end{eqnarray}
where $\mu_0$ stands for an arbitrary initial scale and $a_s=\alpha_s/\pi$. Under the conventional scale setting, the renormalization scale shall be fixed to $\mu_0$; while for a certain scale setting approach, the renormalization scale shall be varied from $\mu_0$ to a certain degree.

The quantity $R_n(Q,\mu_0)$ has been calculated up to four-loop levels under the $\overline{\rm MS}$-scheme~\cite{Ralphas1,Ralphas2}, whose coefficients for $\mu_0=Q$ read
\begin{eqnarray}
{\cal C}_0 &=& 1, \nonumber\\
{\cal C}_1 &=& 1.9857 - 0.1152 n_f, \nonumber\\
{\cal C}_2 &=& -6.63694 - 1.20013 n_f - 0.00518 n_f^2 -1.240 \eta , \nonumber\\
{\cal C}_3 &=& -156.61 + 18.77 n_f - 0.7974 n_f^2 + 0.0215 n_f^3, \nonumber
\end{eqnarray}
where $\eta={\left(\sum_q e_q\right)^2}/{\left(3\sum_q e_q^2\right)}$, $n_f$ and $e_q$ stand for the number and electric charge of the active flavors. Because of the factorial-growth of renormalon terms, the magnitude of the coefficient ${\cal C}_i$ generally grows with the increment of QCD loops, providing the dominant source for lessening the convergence of pQCD series. By applying the PMS, we shall show such kind of factorial growth can be softened to a certain degree.

To do the numerical calculation, the QCD parameter $\Lambda_{\overline{\rm{MS}}}$ is fixed by using $\alpha_s(M_Z)=0.1185\pm0.0006$~\cite{pdg}. For self-consistency, the $\Lambda_{\overline{\rm{MS}}}$ for $R_n$ shall be determined by using $(n + 1)_{\rm th}$-loop $\alpha_s$-running determined from the RG-equation (\ref{beta}). For example, we obtain $\Lambda_{\overline{\rm{MS}}}^{(n_f=5)} =214$ MeV for $R_3$ by using four-loop $\alpha_s$-running. Under the conventional scale setting, the renormalization scale shall be fixed to $\mu_0$; while for the PMS, the renormalization scale shall be the optimal one determined from local RG-invariance. In the following discussions, if not specially stated, we shall take $\mu_0=Q$.

\begin{table}[htb]
\centering
\begin{tabular}{cccc}
\hline
                       & ~~$n_f$=3~~  & ~~$n_f$=4~~  & ~~$n_f$=5~~  \\
\hline
${\cal C}_1$           & 1.6401   & 1.5249   & 1.4097   \\
${\cal C}_2$           & -10.284  & -11.6857 & -12.8047 \\
${\cal C}_3$           & -106.896 & -92.9124 & -80.0075 \\
\hline
~~${\cal C}_1^{\rm PMS}~~ $ & -0.458  & -0.1105  & 0.0479   \\
${\cal C}_2^{\rm PMS}$ & -1.1361  & 0.2103   & 1.3075   \\
${\cal C}_3^{\rm PMS}$ & 32.2133  & 24.9881  & 16.4108  \\
\hline
\end{tabular}
\caption{Coefficients for the perturbative expansion of $R_3(Q)$ before and after the PMS scale setting, where we have set $Q=1.2$ GeV for $n_f$=3, $Q=3$ GeV for $n_f$=4, and $Q=31.6$ GeV for $n_f$=5. } \label{Reecoe}
\end{table}

The coefficients ${\cal C}_1$, ${\cal C}_2$ and ${\cal C}_3$ before and after the PMS scale setting for various flavor numbers, i.e. $n_f=3$, 4, and 5, are presented in Table \ref{Reecoe}. Three typical scales, $Q=1.2$ GeV, $3$ GeV, and $31.6$ GeV, are adopted for various flavor numbers. After applying the PMS, the magnitude of the coefficients ${\cal C}_i$ become smaller than those under the conventional scale setting, indicating that the divergent renormalon terms have been suppressed.

\begin{table}[tb]
\centering
\begin{tabular}{cccc}
\hline
                         & ~~$n_f=3$~~ & ~~$n_f=4$~~ & ~~$n_f=5$~~ \\ \hline
~~$\tilde{a}_s^{(1)}({\rm Conv.})$~~ & 0.1414  & 0.0823  & 0.0450  \\
$\tilde{a}_s^{(2)}({\rm Conv.})$ & 0.1320  & 0.0814  & 0.0450  \\
$\tilde{a}_s^{(3)}({\rm Conv.})$ & 0.1370  & 0.0820  & 0.0450  \\ \hline
$\tilde{a}_s^{(1)}({\rm PMS})$   & 0.2156  & 0.1052  & 0.0504  \\
$\tilde{a}_s^{(2)}({\rm PMS})$   & 0.1265  & 0.0832  & 0.0464  \\
$\tilde{a}_s^{(3)}({\rm PMS})$   & 0.1212  & 0.0819  & 0.0461  \\ \hline
\end{tabular}
\caption{The effective coupling $\tilde{a}_s^{(n)}$ under the conventional (Conv.) and the PMS scale settings, where $n=1$, $2$, and $3$, respectively. Here we have set $Q=1.2$ GeV for $n_f$=3, $Q=3$ GeV for $n_f$=4, and $Q=31.6$ GeV for $n_f$=5. } \label{Reecou}
\end{table}

As for conventional scale setting, one usually takes the same renormalization scale for pQCD predictions up to any perturbative order. Then, under conventional scale setting, the effective coupling $\tilde{a}_s^{(n)} \equiv {a}_s^{(n)}$, and the slight differences among $\tilde{a}_s^{(n)}$ with various $n$ are directly caused by the conventional $\alpha_s$-behavior up to $(n+1)$-loops. After applying the PMS, we shall have different effective/optimal coupling $\tilde{a}_s^{(n)}({\rm PMS})$ for each $R_n$. The effective coupling $\tilde{a}_s^{(n)}$ for $R_n$ under those two scale settings are shown in Table \ref{Reecou}, where $n=1$, $2$, and $3$, respectively. To determine the PMS effective coupling, one does not need to know the value of $\Lambda_{\rm QCD}$, thus the uncertainties from $\Lambda_{\rm QCD}$ are eliminated \footnote{This property has been adopted for dealing with the coupling constant's fixed-point behavior at the low-energy region~\cite{fixed,fixed2}. A detailed PMS analysis on physical observables at the low-energy region in comparison with those of PMC and conventional scale settings is in preparation~\cite{lowenergy}.}. It is noted that the PMS effective coupling $\tilde{a}_s^{(n)}({\rm PMS})$ becomes smaller for a larger $n$, i.e. $\tilde{a}_s^{(1)}({\rm PMS}) > \tilde{a}_s^{(2)}({\rm PMS}) > \tilde{a}_s^{(3)}({\rm PMS})$. This agrees with the previous observation of Ref.~\cite{PMSnpb} and is consistent with the ``induced convergence''~\cite{induce0}.

Next, we turn to numerical analysis of $R_n$ under the PMS and conventional scale settings. For the purpose, we fix $Q=31.6$ GeV, at which the $R$-ratio has been measured~\cite{experimentRee}.

\begin{table}[htb]
\centering
\begin{tabular}{ c c c c c c c c}
\hline
      & LO      & NLO     & N$^2$LO  & N$^3$LO  & $total$ \\ \hline
~Conv.~ & ~0.04499~ & ~0.00285~ & ~-0.00117~ & ~-0.00033~ & ~0.04635~ \\
PMS   & 0.04608 & 0.00010 & 0.00013  & 0.00007  & 0.04638 \\
\hline
\end{tabular}
\caption{The LO, NLO, N$^2$LO and N$^3$LO loop contributions for the approximant $R_3$ under the conventional (Conv.) and the PMS scale settings. The $total$-column stands for the sum of all those loop corrections. $Q=31.6$ GeV. } \label{Rorder}
\end{table}

Given a perturbative series, it is important to know how well it behaves; {\it i.e.,} how much each loop term contributes. In Table \ref{Rorder}, we present the numerical results for the LO, NLO, N$^2$LO and N$^3$LO loop contributions to $R_3$ separately, in which the results for the conventional and the PMS scale settings are given. After applying the PMS, the magnitudes of the NLO, N$^2$LO and N$^3$LO loop-terms become much smaller than the corresponding ones under the conventional scale setting. This is due to the combined effect of the suppression of renormalon terms and the ``induced convergence". However, this does not mean a more convergent pQCD series can be achieved. As shown by Table \ref{Rorder}, the pQCD series under the conventional scale setting has a standard perturbative convergence
\begin{displaymath}
|R^{\rm LO}_{3,{\rm Conv.}}| \gg |R^{\rm NLO}_{3,{\rm Conv.}}| > |R^{\rm N^2LO}_{3,{\rm Conv.}}| > |R^{\rm N^3LO}_{3,{\rm Conv.}}|,
\end{displaymath}
which is mainly caused by $\alpha_s$-power suppression. On the other hand, the PMS prediction shows a quite different perturbation series, {\it i.e.,}
\begin{displaymath}
R^{\rm LO}_{3,{\rm PMS}} \gg R^{\rm NLO}_{3,{\rm PMS}} \sim R^{\rm N^2LO}_{3,{\rm PMS}} \sim R^{\rm N^3LO}_{3,{\rm PMS}}
\end{displaymath}
with $R^{\rm N^{2}LO}_{3,{\rm PMS}} > R^{\rm NLO}_{3,{\rm PMS}}$. The PMS prediction is determined by local RG-invariance, thus its goal is to achieve the steady behavior of a perturbative series other than to improve its pQCD convergence. For example, the LO term $R^{\rm LO}_{3,{\rm PMS}}$ provides over $99\%$ contributions to the PMS series, and the PMS prediction quickly approaches its steady behavior. However, its pQCD convergence can only be an accidental or it shall not show pQCD convergence at all.

\begin{table}[htb]
\centering
\begin{tabular}[b]{cccccccc}
\hline
      & $R_1$   & $R_2$   & $R_3$   & $\kappa_1$ & $\kappa_2$ & $\kappa_3$ \\ \hline
Conv. & 0.04786 & 0.04666 & 0.04635 & $7.44\%$   & $-2.50\%$  & $-0.66\%$ \\
PMS   & 0.04889 & 0.04644 & 0.04638 & $9.76\%$   & $-5.00\%$  & $-0.14\%$ \\
\hline
\end{tabular}
\caption{Numerical results for $R_n$ and $\kappa_n$ with various QCD loop corrections under the conventional (Conv.) and PMS scale settings. The value of $R_0=0.04454$ is the same for both scale settings. $Q=31.6$ GeV. } \label{RnPMS}
\end{table}

To show to what degree a low-order prediction can be improved by a high-order one, we define a ratio
\begin{displaymath}
\kappa_{n}=\frac{R_n-R_{n-1}}{R_{n-1}}.~~~n=(1,2,3)
\end{displaymath}
To be a ``convergent and accurate'' $(n+1)$-loop pQCD prediction, one would think that the magnitude of $\kappa_n$ should be small enough and also be smaller than $\kappa_{(n-1)}$. Numerical results for $R_n$ and $\kappa_n $ up to four-loop level before and after the PMS scale setting are presented in Table \ref{RnPMS}. It shows that both conventional and PMS scale settings can give acceptable predictions when more high-order corrections have been taken into consideration. Up to four-loop level, the absolute values of $\kappa_3$ for the conventional and PMS scale settings are smaller than $1\%$, indicating that the pQCD predictions for this case are convergent and accurate enough; {\it i.e.,} the four-loop prediction $R_3$ are very close to the ``true" value of the physical observable $R$. Following the trends of those predictions, we can expect that the physical value of $R$ could be around $0.04635$.

Previously, there was a doubt casted on the usefulness of PMS~\cite{antisetting} for that it gives larger $\kappa_1$ and $\kappa_2$ than the conventional scale setting does. However the absolute value of PMS $\kappa_3$ is smaller than its counterpart of the conventional scale setting by about three times. This indicates that a larger PMS $\kappa_1$ and $\kappa_2$ only reflect the importance of ${\rm N}^3$-LO correction for PMS to achieving a better prediction than the conventional scale setting. Thus the available ${\rm N}^3$-LO correction helps us to clarify such kind of doubts on PMS.

\begin{figure}[htb]
\centering
\includegraphics[width=0.45 \textwidth]{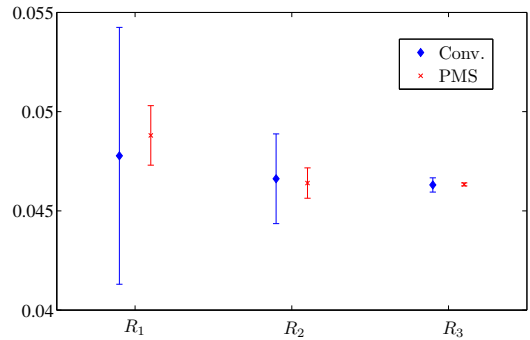}
\caption{Results for $R_n$ ($n=1,2,3$) together with their error estimates $\left(\pm |\tilde{\cal C}_{n}\tilde{a}^{n+1}_s|_{\rm MAX} \right)$. The diamonds and the crosses are for conventional (Conv.) and PMS scale settings, respectively. $Q=31.6\rm{GeV}$. }\label{error}
\end{figure}

\begin{figure}[tb]
\centering
\includegraphics[width=0.45 \textwidth]{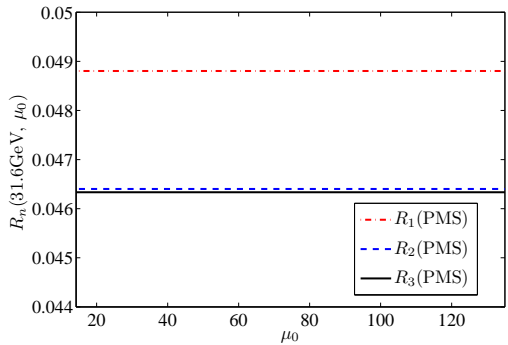}
\caption{The pQCD prediction $R_{n}^{\rm PMS}(Q=31.6 {\rm GeV},\mu_{0})$ up to four-loop level versus the initial scale $\mu_0$. The dash-dot, the dashed and the solid lines are for, $R_1$, $R_2$ and $R_3$, respectively. } \label{RsmuPMS}
\end{figure}

It is helpful to find a way to predict ``unknown" high-order pQCD corrections. Conventionally, this is done by varying the renormalization scale over a certain range, e.g. $\mu_0\in[Q/2, 2Q]$. This conventional error estimate is not reliable, since it only partly estimates the high-order non-conformal contribution but not the more important conformal one~\cite{pmcreview}. It is no reason to choose $1/2$ or $2$ to discuss the error, why not $3$ times or others ? Moreover, for a scale setting such as PMS or PMC, it is unreasonable to simply vary their optimal scales via a similar way to predict ``unknown" high-order pQCD corrections, since this way breaks the RG-invariance and leads to unreliable results. As a conservative prediction, one can take the perturbative uncertainty to be one of the last known order~\cite{pmcma}, i.e. the ``unknown" high-order pQCD correction is taken as $\left(\pm |{\cal C}_{n} \tilde{a}^{n+1}_s|_{\rm MAX} \right)$ for a $(n+1)$-loop prediction of $R_n$, where $|{\cal C}_{n} a^{n+1}_s|$ is calculated by varying $\mu_0\in[Q/2,2Q]$ \footnote{As shown by the latter Fig.(\ref{RsmuPMS}) the PMS prediction is independent to the choice of $\mu_0$, thus such choice of usual scale range only leads to a smaller conventional scale error.}, and the symbol ``MAX'' stands for the maximum $|{\cal C}_{n} a^{n+1}_s|$ within this scale region. The error estimates for conventional and PMS scale settings are displayed in Fig.(\ref{error}). It shows that the PMS errors are smaller than those under the conventional scale setting, which tend to shrink more rapidly with the increment of pQCD order. It is noted that the PMS $R_2$ and $R_3$ lie well outside the error estimation of $R_1$. Thus the PMS prediction on $R_1$ along is not able to predict correct high-order contributions. Such an improper PMS prediction on $R_1$ also explains why PMS $\kappa_1$ and $\kappa_2$ are so large. However by including more high-order contributions, the PMS works better and gives more reliable predictions.

\begin{figure}[tb]
\centering
\includegraphics[width=0.45 \textwidth]{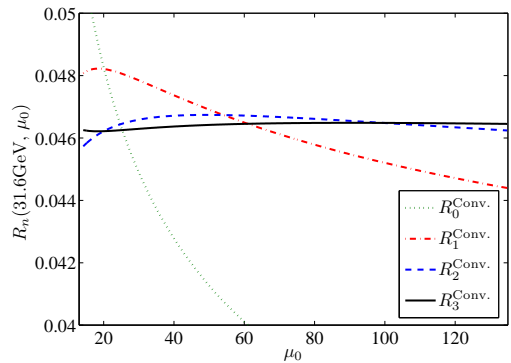}
\caption{The pQCD prediction $R_{n}^{\rm Conv.}(Q=31.6 {\rm GeV},\mu_{0})$ up to four-loop level versus the initial scale $\mu_0$. The dotted, the dash-dot, the dashed and the solid lines are for $R_0$, $R_1$, $R_2$ and $R_3$, respectively. } \label{Rsmu1}
\end{figure}

Finally, we discuss the scale dependence of $R_n$ under different scale settings. We present the scale dependence of $R_n^{\rm Conv.}(31.6{\rm GeV},\mu_0)$ up to four-loop level under the conventional scale setting in Fig.(\ref{Rsmu1}). The LO and NLO estimations, $R_0$ and $R_1$, depend heavily on $\mu_0$. When more high-order corrections have been taken into account, the scale dependence becomes weaker. This agrees with the conventional wisdom that by computing high-order enough correction, one may get scale independent predictions. However, not all quantities in pQCD can be calculated to ``accurate enough'' high orders due to the complexity of high-loop QCD calculations. As a comparison, we present the scale dependence of $R_n^{\rm PMS}(31.6{\rm GeV})$ under the PMS scale setting in Fig.(\ref{RsmuPMS}). It shows that the PMS does eliminate the initial scale dependence even for low fixed-order pQCD predictions, which is consistent with our previous conclusions drawn from the properties of RG-invariants.

\subsection{$R_\tau$ up to four-loop level}
\label{sec:Rtau}

The ratio for $\tau$-lepton decays into hadrons is defined as
\begin{eqnarray}
R_{\tau}=\frac{\Gamma(\tau\rightarrow\nu_\tau+\rm{hadrons})} {\Gamma(\tau\rightarrow\nu_\tau+e^-\bar\nu_e)},
\end{eqnarray}
which provides another fundamental test of pQCD and it can be calculated from $R_{e^+e^-}$~\cite{Rtau1,Rtau2}:
\begin{displaymath}
R_{\tau}(M_\tau)=2\int_0^{M_{\tau}^2}\frac{ds}{M_{\tau}^2} \left(1-\frac{s}{M_{\tau}^2}\right)^2\left(1+\frac{2s}{M_{\tau}^2}\right) \tilde{R}_{e^+e^-}(\sqrt{s}).
\end{displaymath}
Here $M_{\tau}=1.777$ GeV~\cite{pdg} is the $\tau$-lepton mass, $s$ stands for the squared invariant mass of hadrons, and $\tilde{R}_{e^+e^-}(\sqrt{s})$ can be obtained from $R_{e^+e^-}$ by replacing $3\sum_q e_q^2$ with $3(|V_{ud}|^2+|V_{us}|^2)\approx 3$.

After doing the integration over $s$ and putting the explicit scale dependence into the expression, we can rewrite $R_\tau$ as
\begin{eqnarray}
R_{\tau}(M_{\tau},\mu_0)=3(|V_{ud}|^2+|V_{us}|^2) (1+r^{\tau}_{n}(M_{\tau},\mu_0)),
\end{eqnarray}
where the perturbative approximant
\begin{eqnarray}
r^{\tau}_{n}(M_{\tau},\mu_0)&=&\sum_{i=0}^{n} {\cal C}'_{i}(M_{\tau},\mu_0) a_s^{i+1}(\mu_0). \label{rtau}
\end{eqnarray}
$\mu_0$ stands for initial renormalization scale. At $\mu_0=M_{\tau}$, the coefficients of $R_{\tau}$ under the $\overline{\rm MS}$-scheme up to four-loop level can be written as~\cite{Ralphas1}
\begin{eqnarray}
{\cal C}'_0&=&1, \nonumber\\
{\cal C}'_1&=&6.3399-0.3791 n_f, \nonumber\\
{\cal C}'_2&=&48.5831-7.87865 n_f+0.15786 n_f^2, \nonumber\\
{\cal C}'_2&=&401.54-109.449 n_f+6.18148 n_f^2-0.06366 n_f^3. \nonumber
\end{eqnarray}

\begin{figure}[htb]
\centering
\includegraphics[width=0.5 \textwidth]{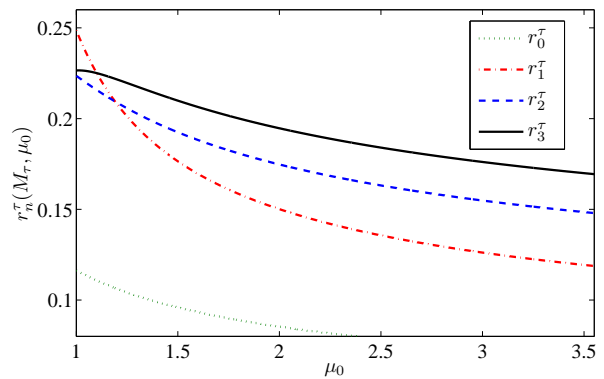}
\caption{The pQCD prediction $r^{\tau}_n(M_{\tau},\mu_0)$ up to four-loop level versus the initial scale $\mu_0$ under the conventional scale setting. The dotted, the dash-dot, the dashed and the solid lines are for $r^{\tau}_0$, $r^{\tau}_1$, $r^{\tau}_2$ and $r^{\tau}_3$, respectively. }\label{Rtaumu1}
\end{figure}

We start from the (initial) scale dependence of $r^{\tau}_n(M_{\tau})$. The results for $r^{\tau}_n$ under the conventional scale setting are put in Fig.(\ref{Rtaumu1}). It is found that the approximant $r^{\tau}_n$ strongly depends on $\mu_0$ even for the four-loop prediction. This indicates that we need even more loop terms to make the final prediction accurate enough. On the other hand, after applying the PMS, we get the same initial scale independence at any orders as that of Fig.(\ref{RsmuPMS}). In the following, we shall take $\mu_0=M_{\tau}$ to do our discussions.

\begin{table}[htb]
\centering
\begin{tabular}{cccc}
\hline
      & ~~${\cal C}'_1$~~ & ~~${\cal C}'_2$~~ & ~~${\cal C}'_3$~~  \\ \hline
~Conv.~ & 5.2023       & 26.3659      & 127.079   \\
PMS   & 0.3906       & 1.2380       & -6.1747 \\ \hline
\end{tabular}
\caption{Coefficients for the perturbative expansion of $r^{\tau}_3$ before and after the PMS scale setting. $\mu_0=M_{\tau}$. } \label{Rtaucoe}
\end{table}

\begin{table}[htb]
\begin{tabular}{cccc}
\hline
      & ~~$\tilde{a}_s^{(1)}$~~ & ~~$\tilde{a}_s^{(2)}$~~ & ~~$\tilde{a}_s^{(3)}$~~ \\ \hline
~Conv.~ & 0.1042      & 0.1015      & 0.1032      \\
PMS   & 0.4733      & 0.1963      & 0.1994      \\ \hline
\end{tabular}
\caption{The effective couplings $\tilde{a}_s^{(n)}$ for $r^{\tau}_n$ under the conventional (Conv.) and PMS scale settings. $\mu_0=M_{\tau}$. } \label{Rtaucou}
\end{table}

The coefficients ${\cal C}'_n$ before and after the PMS scale setting are presented in Table \ref{Rtaucoe}. Again the factorial renormalon growth of ${\cal C}'_n$ has been suppressed. The effective couplings $\tilde{a}_s^{(n)}$ for $r^{\tau}_n$ under the conventional and PMS scale settings are presented in Table \ref{Rtaucou}. In different to the case of $R_n$, there is no ``induced convergence'' for $r^{\tau}_n$, i.e. $\tilde{a}_s^{(1)} > \tilde{a}_s^{(2)} \sim \tilde{a}_s^{(3)}$. Thus the ``induced convergence'' can only be an approximate property of PMS.

\begin{table}[htb]
\begin{tabular}{cccccc}
\hline
      & LO      & NLO     & N$^2$LO & N$^3$LO  & $total$ \\ \hline
Conv. & 0.10320 & 0.05541 & 0.02898 & 0.01441  & 0.20200 \\
PMS   & 0.19935 & 0.01552 & 0.00981 & -0.00975 & 0.21493 \\ \hline
\end{tabular}
\caption{The LO, NLO, N$^2$LO and N$^3$LO loop contributions for the approximant $r^{\tau}_3$ under the conventional (Conv.) and the PMS scale settings. The $total$-column stands for the sum of all those loop corrections. } \label{Rtauorder}
\end{table}

In Table \ref{Rtauorder}, we present numerical results for the LO, NLO, N$^2$LO and N$^3$LO loop contributions to $r^{\tau}_3$ separately, in which the results for conventional and PMS scale settings are presented. The magnitude of ${\rm N}^3$-LO term under the conventional scale is about $7\%$ of $r^{\tau}_3$, which changes down to $\sim4\%$ after applying PMS scale setting. The pQCD series under the conventional scale setting shows a standard perturbative convergence similar to the case of $R_n$. And the PMS prediction also shows a different perturbation series, {\it i.e.,}
\begin{displaymath}
r^{\tau,{\rm LO}}_{3,{\rm PMS}} \gg r^{\tau,{\rm NLO}}_{3,{\rm PMS}} > r^{\tau,{\rm N^2LO}}_{3,{\rm PMS}} \sim \left|r^{\tau,{\rm N^3LO}}_{3,{\rm PMS}}\right|
\end{displaymath}

\begin{table}[htb]
\begin{tabular}{ccccccc}
\hline
      & $r^{\tau}_1$ & $r^{\tau}_2$ & $r^{\tau}_3$ & $\kappa^\tau_1$ & $\kappa^\tau_2$ & $\kappa^\tau_3$ \\ \hline
Conv. & 0.16064      & 0.18255      & 0.20200      & 79.18\%         & 13.64\%         & 10.66\%    \\
PMS   & 0.36514      & 0.19781      & 0.21493      & 307.29\%        & -45.83\%        & 8.66\%     \\ \hline
\end{tabular}
\caption{Numerical results for $r^{\tau}_n$ and $\kappa^\tau_n$ with various QCD loop corrections under the conventional (Conv.) and PMS scale settings. The value of $r^{\tau}_0=0.0897$ is the same for both scale settings. $\mu_0=M_\tau$. } \label{rtautable1}
\end{table}

Numerical results for $r^\tau_n$ and $\kappa^\tau _n$ under conventional and PMS scale settings are presented in Table \ref{rtautable1}. The value of $\kappa^\tau_3$ under both scale settings are around $10\%$, indicating the necessity of calculating more high-order terms before an accurate pQCD prediction on $R_\tau$ can be achieved. $\kappa^\tau_3({\rm PMS})$ is slightly smaller than $\kappa^\tau_3({\rm Conv.})$, thus PMS can lead to relatively better four-loop prediction than the conventional scale setting. The PMS $r^{\tau}_1$ is about $2.2$ times larger than the conventional one, which provides the reason for large $\kappa^\tau_1$ and $\kappa^\tau_2$.

\begin{figure}[htb]
\centering
\includegraphics[width=0.5 \textwidth]{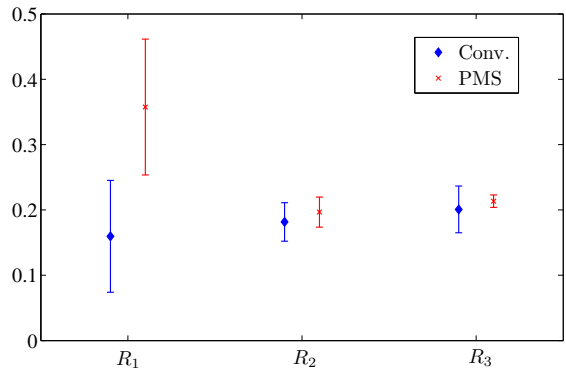}
\caption{Results for $r^{\tau}_n$ ($n=1,2,3$) together with their errors $\left(\pm |\tilde{\cal C}'_{n} \tilde{a}^{n+1}_s|_{\rm MAX} \right)$. The diamonds and the crosses are for conventional scale setting (Conv.) and the PMS.}\label{errortau}
\end{figure}

Results for $r^{\tau}_n$ ($n=1,2,3$) together with their error estimates, i.e. the predicted unknown high-order contributions $\left(\pm |\tilde{\cal C}'_{n} \tilde{a}^{n+1}_s|_{\rm MAX} \right)$, are presented in Fig.(\ref{errortau}). Similar to $R_{e^+e^-}$ case, $r^{\tau}_{2,3}$ are outside the prediction of $r^{\tau}_1$. The PMS $r^{\tau}_1$ is even outside the conventional prediction of $r^{\tau}_1$ with large errors. Thus, the PMS prediction on $r^{\tau}_1$ along is not able to predict correct high-order contributions. But the PMS provides smaller errors for $r^{\tau}_2$ and $r^{\tau}_3$ than those given by the conventional method, and the PMS errors shrink quickly when more loop corrections are included. Using the four-loop prediction, we obtain
\begin{eqnarray}
R_{\tau}(M,\mu_0)|_{\rm{Conv.}} &=& 3.606\pm0.111,\\
R_{\tau}(M,\mu_0)|_{\rm{PMS}} &=& 3.645\pm0.029.
\end{eqnarray}
where the errors are predicted high-order contributions for $\mu_0\in[M/2,2M]$. Both of them are consistent with the OPAL measurement~\cite{Rtauexp}, $R_{\tau}=3.593\pm0.008$. These values strongly depends on the choice of $\Lambda^{n_f=3}_{\rm QCD}$. Inversely, by using the OPAL data on $R_\tau$ and following the approach suggested in Ref.\cite{pmcma}, we predict $\Lambda^{n_f=3}_{\rm Conv.}$=$340^{+4}_{-5}$ MeV and $\Lambda^{n_f=3}_{\rm PMS}$=$323^{+4}_{-4}$ MeV.

\subsection{$\Gamma(H\rightarrow b\bar{b})$ up to four-loop level}
\label{sec:higgsbb}

The decay width for Higgs decaying into a $b\bar{b}$ pair can be written as
\begin{eqnarray}
\Gamma(H\to b\bar{b})=\frac{3G_{F}M_{H}m_{b}^{2}(M_{H})} {4\sqrt{2}\pi}(1+\tilde{R}_n),
\end{eqnarray}
where $G_{F}$ is the Fermi constant, $M_{H}$ is the mass of Higgs Boson, $m_{b}(M_{H})$ is the $b$-quark $\overline{\rm MS}$ running mass, and up to $(n+1)$-loop level, we have
\begin{displaymath}
\tilde{R}_n(M_H,\mu_0)=\sum_{i=0}^{n}\tilde{\cal C}^{\prime\prime}_i(M_H,\mu_0) a^{i+1}_s(\mu_0),
\end{displaymath}
where $\mu_0$ stands for an arbitrary initial scale. The QCD corrections for the decay width $\Gamma(H\to b\bar{b})$ have been calculated up to four-loop level, cf. Refs.\cite{cor9,cor10,cor7,cor8}; For $\mu_0=M_H$, the four-loop $\tilde{R}_3$ reads~\cite{cor9}
\begin{widetext}
\begin{eqnarray}
\tilde{R}_3 &=&5.6667 a_s(M_{H})+ (35.94-1.359n_f) a_s^2(M_{H})+ (164.14-25.77 n_f+0.259n_f^2) a_s^3(M_{H})\nonumber\\
&& + (39.34 -220.9n_f+9.685 n_f^2-0.0205 n_f^3) a_s^4(M_{H}).
\end{eqnarray}
\end{widetext}

\begin{figure}[htb]
\centering
\includegraphics[width=0.5 \textwidth]{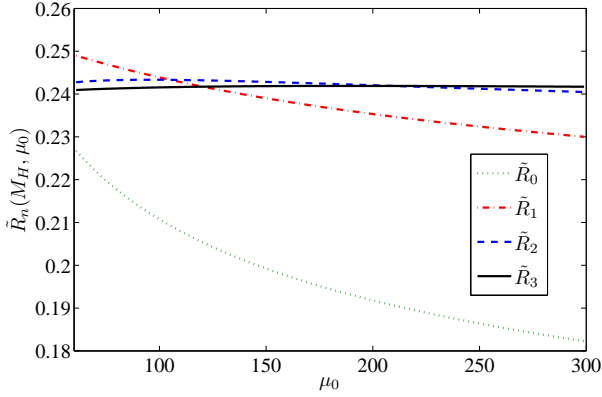}
\caption{The pQCD prediction $\tilde{R}_{n}(M_H,\mu_{0})$ up to four-loop level versus the initial scale $\mu_0$ under the conventional scale setting. The dotted, the dash-dot, the dashed and the solid lines are for $\tilde{R}_0$, $\tilde{R}_1$, $\tilde{R}_2$, and $\tilde{R}_3$, respectively. } \label{Higgsmu1}
\end{figure}

The initial scale dependence of $\tilde{R}_n$ under conventional scale setting is presented in Fig.(\ref{Higgsmu1}). The scale dependence becomes weaker with the increment of high-loop terms, and the four-loop prediction $\tilde{R}_3$ is almost independent to the scale changes. This is the standard properties of pQCD prediction from the conventional scale setting, which however can not weaken the importance of a more proper scale setting. For example, after applying the PMS, we get the same initial scale independence at lower orders as the same as those of Fig.(\ref{RsmuPMS}).

\begin{table}[htb]
\begin{tabular}{cccc}
\hline
      & ~~${\cal C}^{\prime\prime}_1$~~ & ~~${\cal C}^{\prime\prime}_2$~~ & ~~${\cal C}^{\prime\prime}_3$~~ \\ \hline
~Conv.~ & 29.145       & 41.765       & -825.598     \\
PMS   & 0.34376     & 21.2286      & -142.849     \\ \hline
\end{tabular}
\caption{Coefficients for the perturbative expansion of $\tilde{R}_3$ before and after the PMS scale setting. $\mu_0=M_H$. } \label{Hbbcoe}
\end{table}

\begin{table}[htb]
\begin{tabular}{cccc}
\hline
      & ~~$\tilde{a}_s^{(1)}$~~ & ~~$\tilde{a}_s^{(2)}$~~ & ~~$\tilde{a}_s^{(3)}$~~  \\ \hline
~Conv.~ & 0.0360      & 0.0360      & 0.0359 \\
PMS   & 0.0465      & 0.0425      & 0.0423 \\ \hline
\end{tabular}
\caption{The effective couplings $\tilde{a}_s^{(n)}$ for $\tilde{R}_{n}$ under the conventional (Conv.) and PMS scale settings. $\mu_0=M_{H}$. } \label{Hbbcou}
\end{table}

The coefficients ${\cal C}^{\prime\prime}_n$ before and after PMS scale setting in Table \ref{Hbbcoe}. Because of the renormalon term, the absolute value of ${\cal C}^{\prime\prime}_3\sim826$, which changes down to $\sim 143$ by applying PMS. The effective couplings $\tilde{a}_s^{(n)}$ for $\tilde{R}_{n}$ under conventional (Conv.) and PMS scale settings are presented in Table \ref{Hbbcou}. For the present case, the effective couplings $\tilde{a}_s$ for conventional scale setting are almost unchanged, while the PMS ones decreases with the increment of loop-terms.

\begin{table}[htb]
\centering
\begin{tabular}{cccccc}
\hline
      & LO      & NLO     & N$^2$LO & N$^3$LO  & $total$ \\ \hline
Conv. & 0.20371 & 0.03767 & 0.00194 & -0.00138 & 0.24194 \\
PMS   & 0.23967 & 0.00061 & 0.00161 & -0.00046 & 0.24144 \\
\hline
\end{tabular}
\caption{The LO, NLO, N$^2$LO and N$^3$LO loop contributions for the approximant $\tilde{R}_3$ under the conventional (Conv.) and the PMS scale settings. The $total$-column stands for the sum of all those loop corrections. $\mu_0=M_H$. } \label{Higgsorder}
\end{table}

The LO, NLO, N$^2$LO and N$^3$LO loop contributions for the approximant $\tilde{R}_3$ under conventional (Conv.) and PMS scale settings are presented in Table \ref{Higgsorder}. The LO term provides dominant contribution to $\tilde{R}_3$. The magnitude of ${\rm N}^3$-LO term under conventional scale setting provides a smaller $\sim0.6\%$ contribution to $\tilde{R}_3$, which changes down to $0.2\%$ after applying PMS scale setting. The pQCD series under conventional scale setting shows a standard perturbative convergence similar to the case of $R_n$. And the PMS prediction also shows a different perturbation series, {\it i.e.,}
\begin{displaymath}
\tilde{R}^{\rm LO}_{3,{\rm PMS}} \gg \tilde{R}^{\rm NLO}_{3,{\rm PMS}},\tilde{R}^{\rm N^2LO}_{3,{\rm PMS}},\tilde{R}^{\rm N^3LO}_{3,{\rm PMS}}
\end{displaymath}
with $\tilde{R}^{\rm N^2LO}_{3,{\rm PMS}}>\tilde{R}^{\rm NLO}_{3,{\rm PMS}}$.

\begin{table}[htb]
\centering
\begin{tabular}{cccccccccc}
\hline
      & $\tilde{R}_1$ & $\tilde{R}_2$ & $\tilde{R}_3$ & $\tilde{\kappa}_1$ & $\tilde{\kappa}_2$ & $\tilde{\kappa}_3$ \\ \hline
Conv. & 0.24151       & 0.24333       & 0.24194       & $18.28\%$          & $0.75\%$           & $-0.57\%$ \\
PMS   & 0.25621       & 0.24087       & 0.24144       & $25.48\%$          & $-5.99\%$          & $0.24\%$ \\
\hline
\end{tabular}
\caption{Numerical results for $\tilde{R}_n$ and $\tilde{\kappa}_n$ with various QCD loop corrections under the conventional (Conv.) and PMS  scale settings. The value of $\tilde{R}_0=0.20419$ is the same for both scale settings. $\mu_0=M_H$. } \label{tR}
\end{table}

\begin{figure}[htb]
\centering
\includegraphics[width=0.48 \textwidth]{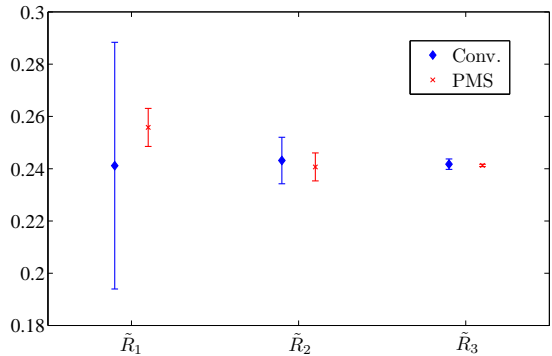}
\caption{Results for $\tilde{R}_n$ ($n=1,2,3$) together with their prediction of unknown high-order contributions $\left(\pm |\tilde{\cal C}^{\prime\prime}_{n} \tilde{a}^{n+1}_s|_{\rm MAX} \right)$ for $H \to b\bar{b}$. The diamonds and the crosses are for conventional (Conv.) and PMS scale settings, respectively.} \label{Higgserror}
\end{figure}

Numerical results for $\tilde{R}_n$ and $\tilde{\kappa}_n$ up to four-loop level are presented in Table \ref{tR}. Results for $\tilde{R}_n$ ($n=1,2,3$) together with their prediction of unknown high-order contributions $\left(\pm |\tilde{\cal C}^{\prime\prime}_{n} \tilde{a}^{n+1}_s|_{\rm MAX} \right)$ are presented in Fig.(\ref{Higgserror}). The four-loop $\tilde R_3$ are nearly the same for conventional and PMS scale settings; while the PMS $\tilde \kappa_3$ is smaller and is more close to its final prediction on the observable $\tilde{R}$. However, the PMS $\tilde{R}_1$ also can not predict the correct high-order contributions, i.e. both $\tilde{R}_2$ and $\tilde{R}_3$ are outside its prediction. Such larger PMS $\tilde{R}_1$ also leads to larger $\tilde \kappa_1$ and $\tilde \kappa_2$. With those $\tilde R_3$ results, we present the decay width of Higgs into a $b\bar{b}$ pair:
\begin{eqnarray}
\Gamma(H\rightarrow b\bar{b})|_{\rm Conv.} &=& 2389.85\pm 3.85 \; {\rm KeV}, \\
\Gamma(H\rightarrow b\bar{b})|_{\rm PMS} &=& 2388.87\pm 0.88 \; {\rm KeV},
\end{eqnarray}
where the errors are predicted unknown high-order contributions for $\mu_0\in[M_H/2,2M_H]$.

\section{A comparison of PMS and PMC}
\label{sec:vs}

The running behavior of the coupling constant is controlled by the RG-equation. In different to the local RG-invariance of PMS, the PMC~\cite{pmc1,pmc2,pmc3,pmc4,pmc5,BMW,BMW2} respects the standard RG-invariance and improves the perturbative series by absorbing all $\beta$-terms governed by RG-equation into the coupling constant. The PMC procedure can be advantageously applied to entire range of perturbatively calculable QCD and Standard Model processes. Recently, many high-order PMC applications have been finished and the PMC works successfully, cf.Refs.\cite{wangsq,app1,app2,app3,app4}. It is helpful to present a detailed comparison of PMS and PMC predictions. For the purpose, we take $R_{e^+e^-}$ as an explicit example.

\begin{table}[htb]
\centering
\begin{tabular}{cccc}
\hline
                       & ~~$n_f$=3~~  & ~~$n_f$=4~~  & ~~$n_f$=5~~  \\ \hline
~${\cal C}_1^{\rm PMC}$~ & 2.14579  & 1.99302  & 1.84024   \\
${\cal C}_2^{\rm PMC}$ & 3.39697  & 1.21574  & -1.00503  \\
${\cal C}_3^{\rm PMC}$ & 6.47103  & -12.8517 & -11.0871  \\ \hline
\end{tabular}
\caption{Coefficients ${\cal C}_{n}^{\rm PMC}$ for the perturbative expansion of $R_3(Q)$ using the PMC scale setting, where we have set $Q=1.2$ GeV for $n_f$=3, $Q=3$ GeV for $n_f$=4, and $Q=31.6$ GeV for $n_f$=5. } \label{Reecoevs}
\end{table}

After applying the PMC, the coefficients ${\cal C}_{n}^{\rm PMC}$ for $R_3$ are presented in Table \ref{Reecoevs}. Comparing with Table \ref{Reecoe}, PMC coefficients are smaller than the conventional ones. PMS also leads to such a suppression, but it can not explain why. PMC shows that such suppression are rightly due to the elimination of renormalon terms.

\begin{table}[htb]
\centering
\begin{tabular}[b]{cccccccc}
\hline
     & $R_1$   & $R_2$   & $R_3$   & $\kappa_1$ & $\kappa_2$ & $\kappa_3$ \\ \hline
PMS  & 0.04889 & 0.04644 & 0.04638 & $9.76\%$   & $-5.00\%$  & $-0.14\%$ \\
PMC  & 0.04767 & 0.04667 & 0.04635 & $7.03\%$   & $-2.09\%$  & $-0.69\%$ \\
\hline
\end{tabular}
\caption{A comparison of $R_n$ and $\kappa_n$ under the PMS and PMC scale settings. The value of $R_0=0.04454$ is the same for both scale settings. $Q=31.6$ GeV and $\mu_0=Q$. } \label{Rnvs}
\end{table}

A comparison of $R_n$ and $\kappa_n$ under PMS and PMC scale settings is presented in Table \ref{Rnvs}. The differences for three-loop $R_2$ is about $0.5\%$, which moves down to about $0.05\%$ for four-loop $R_3$. Both PMS and PMC are based on RG-invariance, it is reasonable that they can give close numerical predictions at higher orders. The values of PMC $\kappa_1$ and $\kappa_2$ are smaller than PMS, indicating a faster steady behavior can be achieved by PMC \footnote{The PMS $\kappa_3$ for $R_{e^+ e^-}$ is accidentally small. We have found that PMC $\kappa_3$ for $R_\tau$ and $\Gamma(H\to b\bar{b})$ are smaller than those of PMS, following the same trends.}.

\begin{table}[htb]
\centering
\begin{tabular}{ c c c c c c c c}
\hline
    & ~~LO~~      & ~~NLO~~     & ~~N$^2$LO~~  & ~~N$^3$LO~~  & ~~$total$~~ \\ \hline
~PMS~ & 0.04608 & 0.00010 & 0.00013  & 0.00007  & 0.04638 \\
PMC & 0.04290 & 0.00351 & -0.00004 & -0.00002 & 0.04635 \\
\hline
\end{tabular}
\caption{The LO, NLO, N$^2$LO and N$^3$LO loop contributions for the approximant $R_3$ under the PMS and PMC scale settings. The $total$-column stands for the sum of all those loop corrections. $Q=31.6$ GeV and $\mu_0=Q$. } \label{Rordervs}
\end{table}

A comparison of PMS and PMC pQCD series is presented in Table \ref{Rordervs}. The PMC pQCD series follows the standard pQCD convergence but is much more convergent than that of conventional scale setting; while, the PMS series also becomes more convergent than the conventional ones, but the series does not show the order-by-order convergence, i.e. $R^{\rm N^{2}LO}_{3,{\rm PMS}} > R^{\rm NLO}_{3,{\rm PMS}}$.

\begin{figure}[htb]
\centering
\includegraphics[width=0.5 \textwidth]{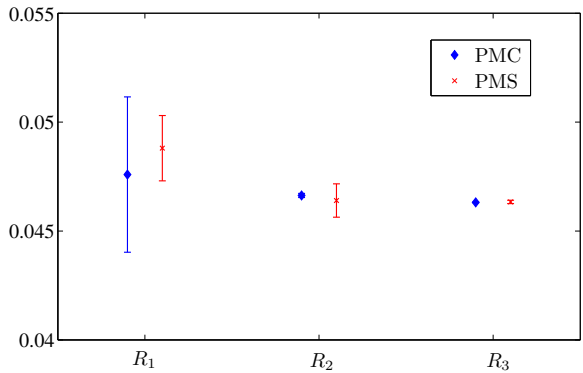}
\caption{A comparison of PMC and PMS predictions for $R_n$ ($n=1,2,3$) together with their predicted unknown high-order contributions $\left(\pm |\tilde{\cal C}_{n} \tilde{a}^{n+1}_s|_{\rm MAX} \right)$. The diamonds and the crosses are for PMC and PMS scale settings, respectively. }\label{errorvs}
\end{figure}

In Fig.(\ref{errorvs}), we present a comparison of PMC and PMS predictions for $R_n$ ($n=1,2,3$) together with their predicted unknown high-order contributions $\left(\pm |\tilde{\cal C}_{n} \tilde{a}^{n+1}_s|_{\rm MAX} \right)$. The large error bar for PMC $R_1$ shows the magnitude of the NLO-conformal terms are large and we need even high-order terms to achieve an accurate prediction. In fact, when we have more $\beta$-terms to fix the PMC scales, the PMC prediction together with its predicted error does become more accurate.

\begin{center}
\begin{figure}
\includegraphics[width=0.5 \textwidth]{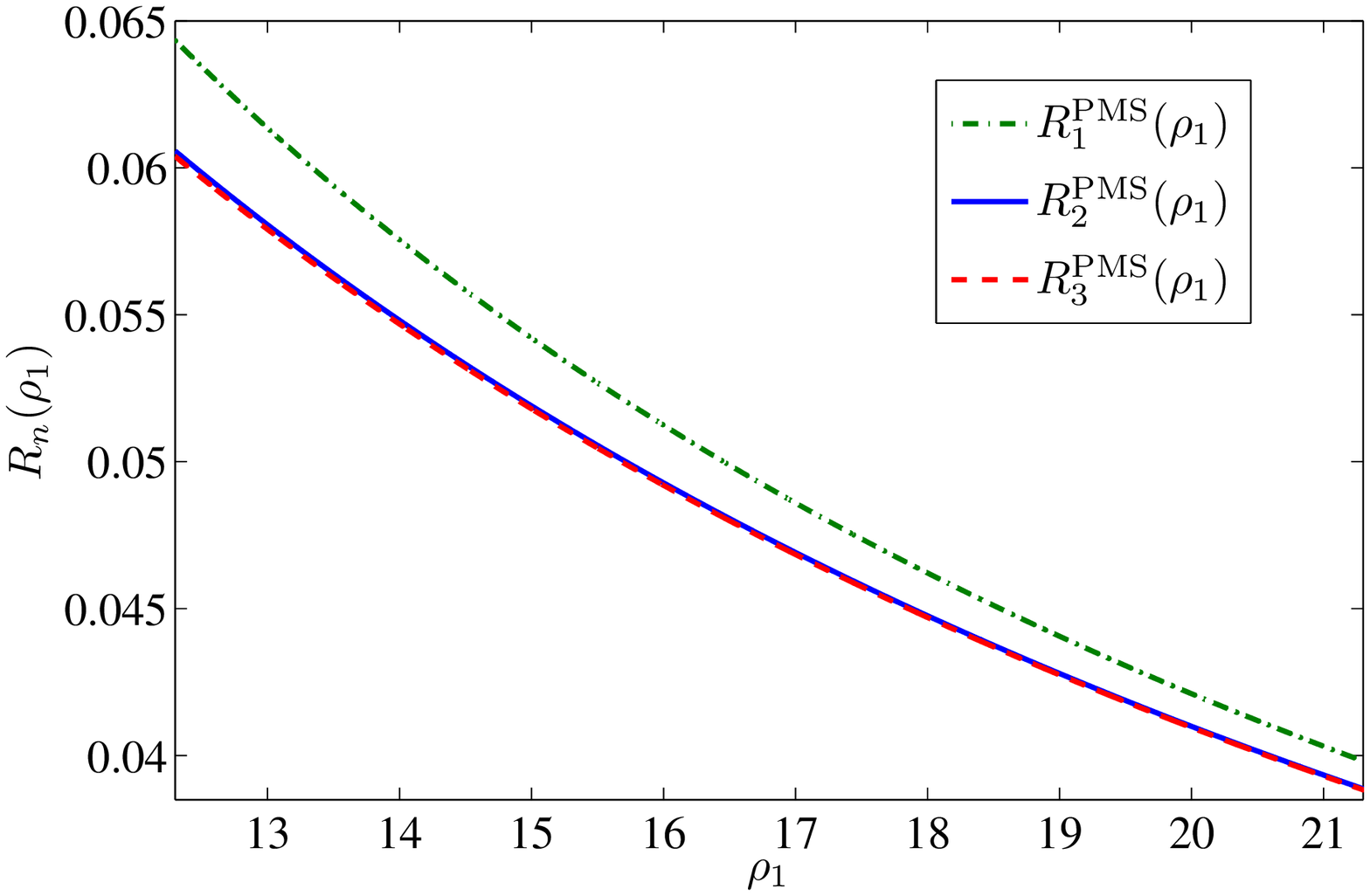}
\includegraphics[width=0.5 \textwidth]{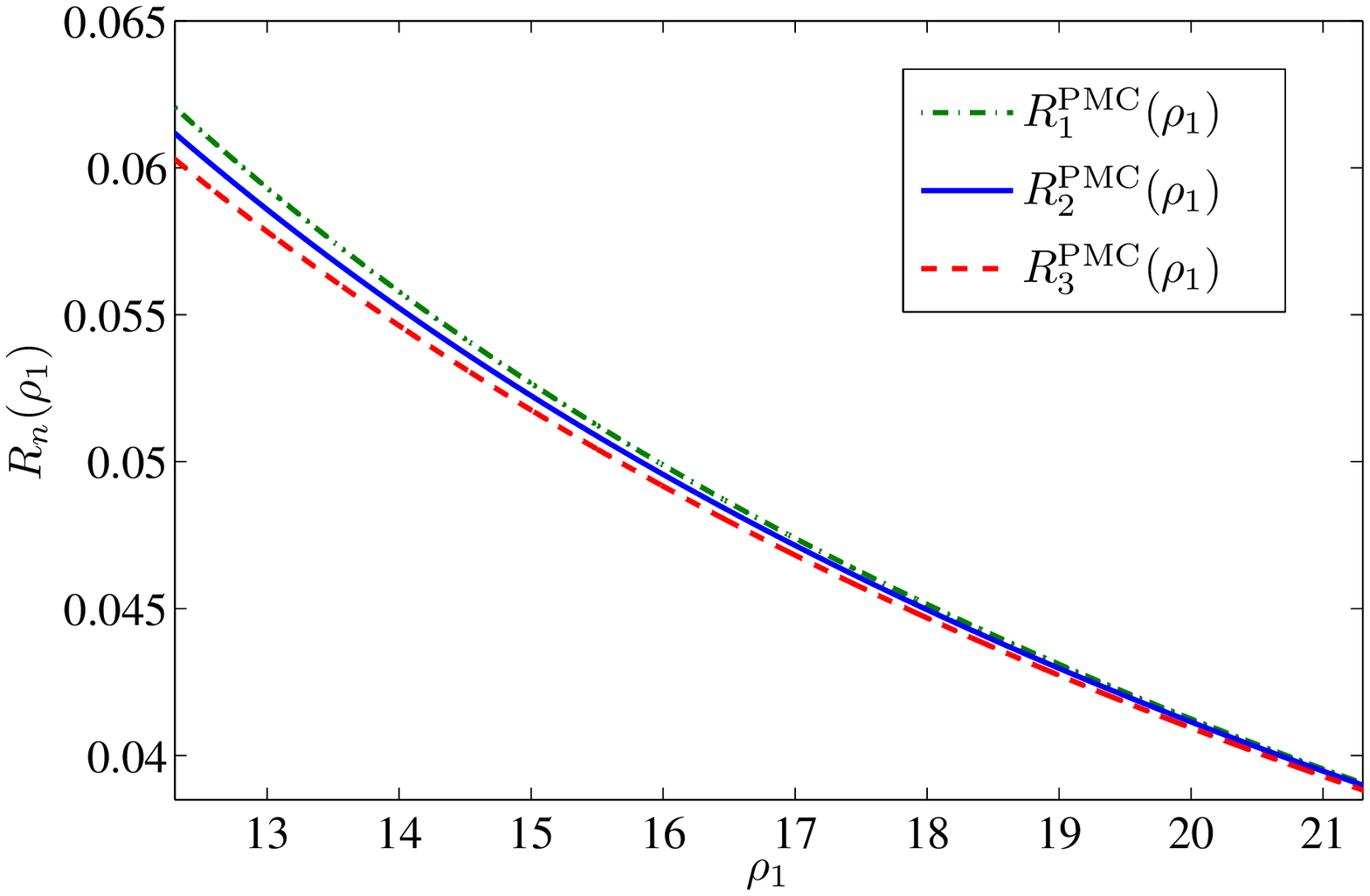}
\caption{The curves of the function $R_n(\rho_1)$ for the PMS and PMC scale settings. $\mu_0=Q$. }\label{rhorun}
\end{figure}
\end{center}

\begin{table}[htb]
\begin{tabular}[b]{ccccc}
\hline
         & ~~$\delta_1$~~ & ~~$\delta_2$~~ & ~~$\delta_3$~~  \\ \hline
~Conv.~  & $2.24\%$     & $0.55\%$     & $0.10\%$  \\
PMC      & $2.66\%$     & $0.61\%$     & $0.07\%$  \\
\hline
\end{tabular}
\caption{The difference $\delta_n$ for $R_n(\rho1)$ between the PMC (or conventional) scale setting and the PMS scale setting.} \label{distance}
\end{table}

We present a comparison of PMS and PMC energy dependence of $R_n(Q)$ in Fig.(\ref{rhorun}), where we have changed the argument to $\rho_1$ such that to avoid the uncertainty from $\Lambda_{\rm{QCD}}$~\cite{PMS2}. The present range $\rho_1\in(12,21)$ corresponds to energy range $9<Q<90$ GeV. There is large difference between $R_1^{\rm PMS}(\rho_1)$ and $R_{n\ge2}^{\rm PMS}(\rho_1)$, which is consistent with previous observation that $R_1^{\rm PMS}(\rho_1)$ along can not predict reasonable unknown high-order contributions. To show the difference of the predicted $R_n(\rho_1)$ under various scale settings more accurately, we define a parameter, $\delta_n$, as
\begin{equation}\label{eq:delta}
\delta_n = \frac{\sum_{\rho_1}|R_n(\rho_1)-R_n^{\rm PMS}(\rho_1)|}{\sum_{\rho_1}|R_n^{\rm PMS}(\rho_1)|}\times 100\%,
\end{equation}
where $\rho_1=12,12.001,12.002,\cdots,21$. Those differences are presented in Table \ref{distance}. The differences of $R_n(\rho_1)$ among different scale settings shall be reduced with more loop corrections being included.

\section{Summary} \label{sec:summary}

To solve the renormalization scheme and renormalization scale ambiguities, one should answer the question of how to set optimal scale systematically for any physical processes up to any orders from some basic principal of QCD theory. As a practical solution, the PMS adopts local RG-invariance (\ref{finite-sca},\ref{finite-sch}) to set the optimal scheme and optimal scale of the process.

Based on the local RG-invariance, we have presented the detailed technology for applying PMS to high-perturbative orders. We have investigated the PMS properties based on three typical physical quantities $R_{e^+e^-}$, $R_\tau$ and $\Gamma(H\rightarrow b\bar{b})$ up to four-loop QCD corrections. Our analysis show that even though the PMS is theoretically unsound, it does provide an effective approach to soften the renormalization scheme and scale ambiguities by including enough higher-order pQCD contributions. More explicitly, our results show that
\begin{itemize}
\item After applying the PMS, the magnitudes of perturbative coefficients become smaller than those under conventional scale setting, indicting the divergent renormalon terms can be suppressed. The PMS effective coupling approximately satisfies the ``induced convergence''. As a combined effect, the magnitudes of NLO and higher-order loop-terms become much smaller than the corresponding ones under conventional scale setting.

\item The goal of PMS is to achieve the steady point of a perturbative series over the renormalization scheme and scale changes. The PMS predictions for those three four-loop examples do show such a steady behavior, i.e. the final PMS predictions are independent of any choice of initial scale, being consistent with one of requirement of basic RG-invariance. Moreover, the LO terms $R^{\rm LO}_{3,{\rm PMS}}$ and $\tilde{R}^{\rm LO}_{3,{\rm PMS}}$ provide $\sim99\%$ contributions, and $r^{\tau,{\rm LO}}_3$ provides $\sim89\%$ contribution to $R_{e^+e^-}$, $\Gamma(H\to b\bar{b})$ and $R_\tau$ series, respectively. However, the PMS have no principal to ensure the pQCD convergence, thus the improved pQCD convergence for some of the high-energy processes could only be an accidental. In fact, all those three four-loop examples do not have standard pQCD convergence, i.e. the magnitudes of their NLO, N$^2$LO and N$^3$LO terms are usually small but at the same order.

\item We have suggested a conservative way to discuss the pQCD predictive power, i.e. to show how unknown high-order terms contribute. It is noted that after PMS scale setting, the N$^2$LO and N$^3$LO estimates are usually outside the predicted errors by using the terms only up to NLO level. Together with other lower-order PMS behaviors, such as the large PMS $\kappa_{1,2}$ for the mentioned processes, we may conclude that PMS can not provide correct lower-order predictions, such as the NLO predictions. In the literature, most of the doubts on PMS are rightly based on lower-order predictions. With more loop corrections being included, the PMS can achieve a more accurate prediction better than that of conventional scale setting.

\end{itemize}

In the paper, we have also presented a comparison of PMS and PMC predictions. In different to PMS, the PMC satisfies standard RG-invariance and follows the RG-equation to fix the running behavior of the coupling constant, thus it is theoretical sound. The PMC predictions have optimal pQCD convergence due to the elimination of renormalon terms. The PMS prediction is independent on the choice of initial scale; while there is residual scale dependence for PMC predictions due to unknown high-order $\beta$-term, however such residual scale dependence is highly suppressed, even for lower-order PMC predictions. In comparison to the conventional and PMS scale settings, the PMC shows a better predictive power, and its predictions quickly approaches the physical value of the observable. Moving to high-order pQCD predictions, the PMS and PMC differences on the pQCD predictions shall be greatly suppressed, e.g. for the case of $R_{e^+e^-}$, the differences change from larger $\sim3\%$ at the NLO level, to be $\sim1\%$ at the N$^2$LO level, and to be less than $0.1\%$ at the N$^3$-LO level.

\hspace{1cm}

\noindent{\bf Acknowledgments}: We thank Stanley J. Brodsky, Hai-Bing Fu, Matin Mojaza, Paul M. Stevenson, Sheng-Quan Wang, and Xu-Chang Zheng for helpful discussions. This work was supported in part by Natural Science Foundation of China under Grant No.11275280, by the Fundamental Research Funds for the Central Universities under Grant No.CQDXWL-2012-Z002.  \\

\appendix

\section{A tricky way to derive the RG-invariants $\rho_n$ at high-orders}

In this appendix, we present a simpler way to drive the RG-invariants $\rho_n$ at high-orders with $n>1$, basing on their properties of RG-invariance.

For convenience, we set $p=1$ in Eq.(\ref{phy}) and redefine the $\beta^{\cal R}$-function as
\begin{eqnarray}
\beta^{\cal R}=\mu^2\frac{\partial}{\partial\mu^2} \left(\frac{\alpha_s(\mu)}{4\pi}\right) =-\sum_{i=0}^{\infty}b_{i}a_s^{i+2},
\end{eqnarray}
where $b_i=(1/4)^{i+2}\beta^{\cal R}_i$ and  $a_s=\alpha_s/\pi$.

A physical observable solitarily defines an effective charge~\cite{FAC1,FAC2,FAC3}, and vice versa. Thus, we can inversely write down the coupling constant $a_s$ as an expression over the approximant $\varrho_n$~\cite{Maxwell}, i.e.
\begin{eqnarray}
a_s(\varrho_n)= \varrho_n  + \sum_{i=1}^{\infty} {r}_i {\varrho_n}^{i+1}  \label{coupling}
\end{eqnarray}
Substituting the $\varrho_n$ expression (\ref{phy}) into Eq.(\ref{coupling}), we obtain
\begin{widetext}
\begin{eqnarray}
a_s &=& a_s \left(1 + {\cal C}_1 a_s+ {\cal C}_2 a_s^2+ {\cal C}_3 a_s^3+\cdots\right) \left[1+r_1 a_s (1 + {\cal C}_1 a_s+ {\cal C}_2 a_s^2+ {\cal C}_3 a_s^3+\cdots) \right. \nonumber\\
&& \left. + r_2 a_s^2 (1 + {\cal C}_1 a_s+ {\cal C}_2 a_s^2+ {\cal C}_3 a_s^3+...)^2+\cdots\right] \\
&=& a_s \left[1+a_s (r_1+{\cal C}_1)+a_s^2 (2 r_1 {\cal C}_1+r_2+{\cal C}_2)+a_s^3 (r_1 {\cal C}_1^2+2 r_1 {\cal C}_2 +3 r_2 {\cal C}_1+r_3+{\cal C}_3 )+\cdots\right],
\end{eqnarray}
\end{widetext}
where the symbol $\cdots$ stands for higher-order terms. The coefficients for $a_s^2$ and higher orders should vanish, which lead to
\begin{eqnarray}
r_1 &=& -{\cal C}_1, \\
r_2 &=& 2 {\cal C}_1^2-{\cal C}_2, \\
r_3 &=& {\cal C}_1^3-3(2{\cal C}_1^2-{\cal C}_2){\cal C}_1+2{\cal C}_1{\cal C}_2-{\cal C}_3 \\
&\vdots& \nonumber
\end{eqnarray}

As a further step, we introduce a new function
\begin{equation}
{\cal R}(Q) = \frac{\partial}{\partial \ln Q^2} \varrho_n(Q) = 4 \beta^{\cal R} \frac{\partial}{\partial a_s(Q)} \varrho_n(Q), \label{rhodefine}
\end{equation}
where $Q$ is the scale at which the observable is measured. Since $\varrho_n$ and $Q$ are physical quantities, ${\cal R}$ can also be regarded as a physical quantity that does not dependent on the renormalization scheme and scale. Eq.(\ref{rhodefine}) can be expanded over $\varrho_n$ in the following form,
\begin{widetext}
\begin{eqnarray}
{\cal R} =&&-\varrho_n^2 [4 b_0+4 \varrho_n (2 r_1 b_0 +b_1 +2 {\cal C}_1 b_0) + 4 \varrho_n^2 (r_1^2 b_0 + 3 r_1 b_1+6 r_1 b_0 {\cal C}_1+2 r_2 b_0+b_2+3 b_0 {\cal C}_2 + 2 b_1 {\cal C}_1)\nonumber \\
&&+4 \varrho_n^3(2 r_1 r_2 b_0+2 r_3 b_0+3 r_1^2 b_1+3 r_2 b_1+4 r_1 b_2+b_3+6 r_1^2 b_0 {\cal C}_1+6 r_2 b_0 {\cal C}_1+8 r_1 b_1 {\cal C}_1+2 b_2 {\cal C}_1\nonumber \\
&&+12 r_1 b_0 {\cal C}_2+3 b_1 {\cal C}_2+4 b_0 {\cal C}_3)+\cdots] \\
=&& -\varrho_n^2 [4 b_0+4 \varrho_n b_1 + 4 \varrho_n^2 (b_2-b_0 {\cal C}_1^2 -b_1 {\cal C}_1+b_0 {\cal C}_2)+4 \varrho_n^3(4 b_0{\cal C}_1^3 -6 b_0{\cal C}_1 {\cal C}_2 +2 b_0{\cal C}_3 + b_1{\cal C}_1^2  -2 b_2 {\cal C}_1 + b_3)+\cdots].
\end{eqnarray}
\end{widetext}

Both $\varrho_n$ and ${\cal R}$ are physical quantities, the expansion coefficients of $\cal R$ over $\varrho_n$ should be RG invariants. Transforming these RG invariant coefficients back into the notation used in the body of the text, we get the RG invariants $\rho_n$ ($n>1$). The first two of them are
\begin{eqnarray}
\rho_2 &=& \frac{\beta_2}{16 \beta_0}-\frac{\beta_1 {\cal C}_1}{4 \beta_0}-{\cal C}_1^2+{\cal C}_2 , \\
\rho_3 &=& \frac{\beta_3}{64 \beta_0}+\frac{\beta_1 {\cal C}_1^2}{4 \beta_0}-\frac{\beta_2 {\cal C}_1}{8 \beta_0}+4 {\cal C}_1^3-6 {\cal C}_2 {\cal C}_1+2 {\cal C}_3 .
\end{eqnarray}

Finally, by replacing $\varrho_n$ to $\varrho_n^{1 \over p}$, we can obtain the RG-invariants for any $p$. The first two of which agree with  Eqs.(\ref{rho21},\ref{rho31}).

\end{document}